\documentclass{moriond}
\usepackage{siunitx}
\usepackage[square,numbers]{natbib}

\def\Journal#1#2#3#4{{#1} {\bf #2}, #3 (#4)}

\def\NIMA{{\em Nucl. Instrum. Methods} A}

\def\PRL{\em Phys. Rev. Lett.}
\def\PRD{{\em Phys. Rev.} D}

\def\be{\begin{equation}}
\def\ee{\end{equation}}
\def\bea{\begin{eqnarray}}
\def\eea{\end{eqnarray}}

\newcommand{\Photo}{\includegraphics[height=35mm]{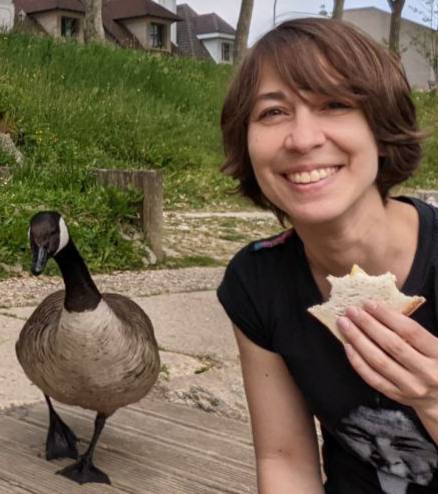}}

\begin{document}
\vspace*{4cm}
\title{Investigating Hadronic Interactions at Ultra-High Energies\\ with the Pierre Auger Observatory}

\author{Isabel Goos \footnote{\href{mailto:goos@apc.in2p3.fr}{goos@apc.in2p3.fr}} on behalf of the Pierre Auger Collaboration \footnote{\href{http://www.auger.org/archive/authors_2022_03.html}{http://www.auger.org/archive/authors\_2022\_03.html}}}

\address{Particles Group, Astroparticle and Cosmology Laboratory (APC), \\10 Rue Alice Domon et Léonie Duquet, 75013 Paris, France}

\maketitle\abstracts{
The development of an extensive air shower depends not only on the nature of the primary ultra-high-energy cosmic ray but also on the properties of the hadronic interactions. For energies above those achievable in human-made accelerators, hadronic interactions are only accessible through the studies of extensive air showers, which can be measured at the Pierre Auger Observatory. With its hybrid detector design, the Pierre Auger Observatory measures both the longitudinal development of showers in the atmosphere and the lateral distribution of particles that arrive at the ground. This way, observables that are sensitive to hadronic interactions at ultra-high energies can be obtained. While the hadronic interaction cross-section can be assessed from the longitudinal profiles, the number of muons and their fluctuations measured with the ground detectors are linked to other physical properties. In addition to these direct studies, we discuss here how measurements of the atmospheric depth of the maximum of air-shower profiles and the characteristics of the muon signal at the ground can be used to test the self-consistency of the post-LHC hadronic models.}

\section{Introduction}

$\quad \;$ The Pierre Auger Observatory is primarily designed to detect cosmic rays with energies above around $10^{18} \: \si{eV}$ via the extensive air showers they generate in the earth's atmosphere. It employs a hybrid technique to measure the properties of the primary cosmic ray \cite{Auger}. On the one hand, a surface detector (SD) consisting of 1600 water-Cherenkov detectors, arranged on an isometric triangular grid with 1500 m spacing, covers an area of 3000 \si{km^2}. This array samples the lateral distribution of the particles of extensive air showers that arrive at the ground. The signal and timing information of triggered stations are used to estimate the primary energy and reconstruct the shower geometry. On the other hand, a fluorescence detector (FD), comprising 27 telescopes and spread between four sites, overlooks the surface detector. These telescopes observe the longitudinal development of extensive air showers through the fluorescence light emitted by de-exciting atmospheric nitrogen molecules. They provide a calorimetric measurement of the primary energy and a complementary description of the shower geometry. Furthermore, they measure the depth $X_\mathrm{max}$ (in \si{g/cm^2}) in the atmosphere where an air shower attains its maximum number of particles. In addition, several instrumental upgrades, an endeavor collectively called AugerPrime, are being deployed in order to disentangle the number of muons $N_{\mu}$ \cite{AugerPrime}. As will be discussed here, this complementary measurement is important to improve the understanding of extensive air showers.

Measuring the energy spectrum of ultra-high energy cosmic rays is of prime importance for understanding the origin and mechanisms of cosmic ray acceleration and propagation. Another open question regarding the energy spectrum is about the origin of the flux suppression observed at the highest energies. Cosmic rays can reach energies up to around $10^{20}$ $\si{eV}$, which corresponds to $\sqrt{s} \sim 200$ $\si{TeV}$ in the center of mass frame. Considering that the highest available energy in a human-made accelerator is of $\sqrt{s} = 13$ $\si{TeV}$ at the LHC, this means that cosmic rays open the unique possibility of studying hadronic interactions at ultra-high energies. Current high-energy hadronic interaction models used to simulate extensive air showers rely on extrapolation of physical quantities towards cosmic ray energies and also phase-space regions inaccessible in man-made experiments. These models are used to compare the observables obtained from simulations with those measured at Auger and to understand the relationship they have with physical parameters.

\begin{figure}
\hspace{1.1cm}
\begin{minipage}{0.32\linewidth}
\centerline{\includegraphics[width=1.6\linewidth]{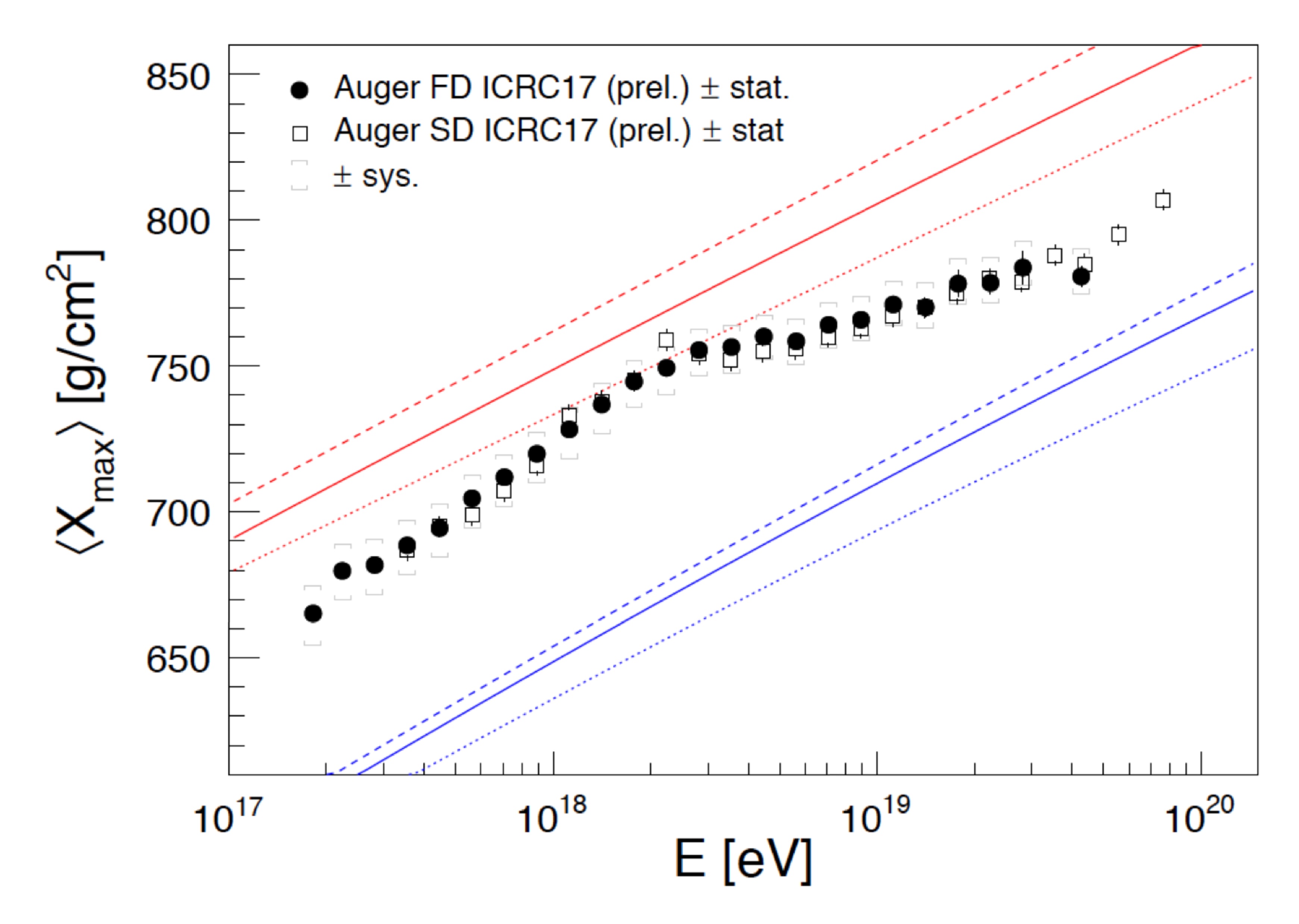}}
\end{minipage}
\hspace{3cm}
\begin{minipage}{0.32\linewidth}
\centerline{\includegraphics[width=1.6\linewidth]{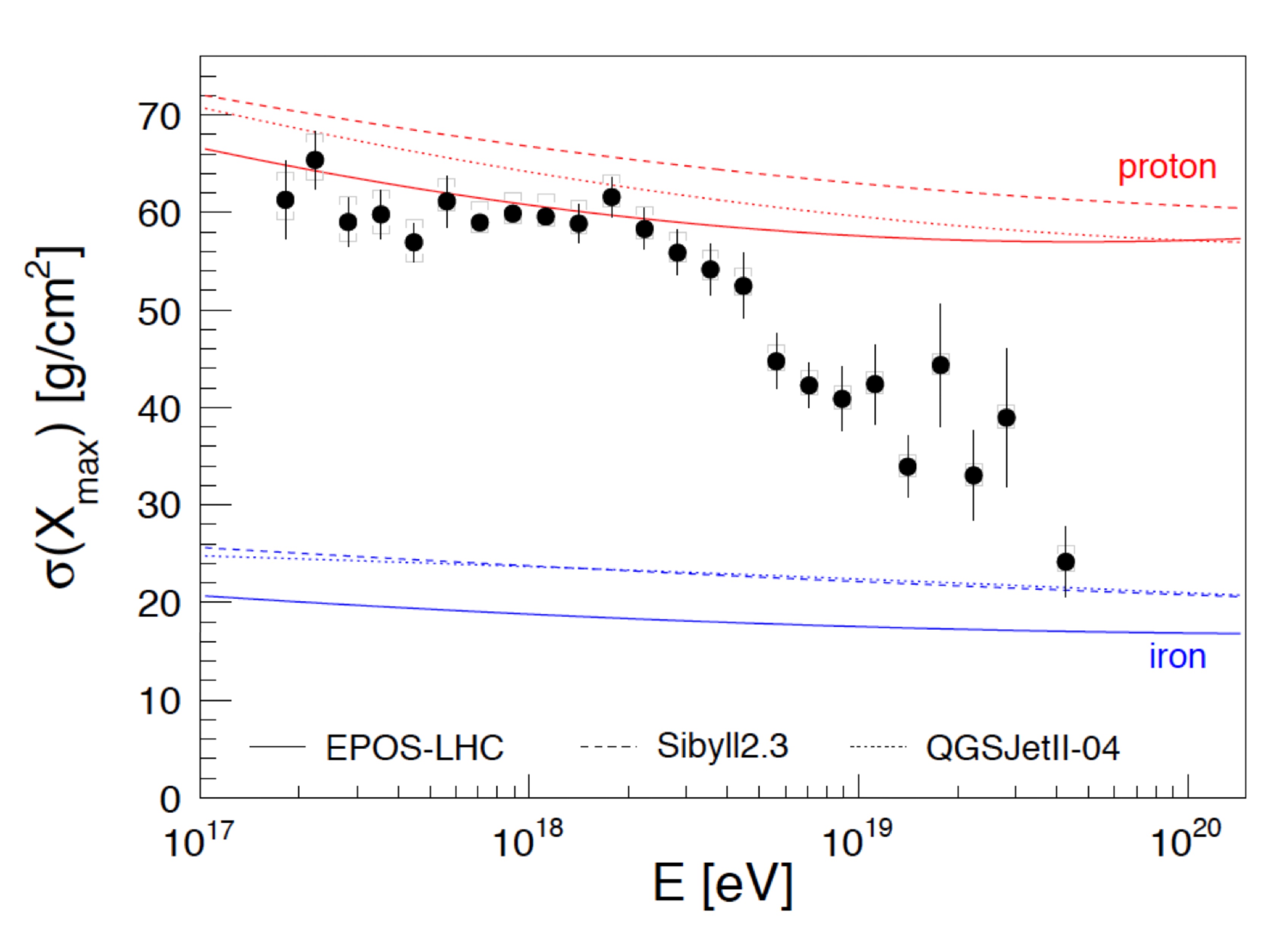}}
\end{minipage}
\caption[]{Energy evolution of the mean of $X_\mathrm{max}$ (left) and its standard deviation (right) as measured at the Pierre Auger Observatory compared to proton and iron shower simulations. Figures from \cite{Xmax}.}
\label{fig:fig1}
\end{figure}

Up to now, none of the available high-energy hadronic interaction models has been able to describe all observables of extensive air showers consistently. Any observed inconsistency hints at a deficiency in the models. Thus, looking for inconsistencies helps understanding where improvement in the models is needed. It is still possible to study the relationship between certain observables and physical parameters, which provides a way of performing direct measurements using cosmic ray observations. It is worth noting that air shower observables are not necessarily independent of one another, which means that performing these studies, considering simultaneously many observables, might render a more complete picture.

\section{Search for inconsistencies}

$\quad \;$ In this section, we present a selection of studies performed to investigate the consistency between measured observables and their values obtained from simulations done using different high-energy interaction models. The generators used in the works reviewed here are EPOS-LHC \cite{Epos}, Sibyll2.3 \cite{Sibyll} and QGSJetII-04 \cite{Qgsjet}. 

\subsection{Depth of maximum development}

$\quad \;$ One of the most robust mass sensitive observables is the depth of the shower maximum $X_\mathrm{max}$. This sensitivity relies on the fact that extensive air showers generated by lighter primaries develop deeper into the atmosphere than those initiated by heavier primaries. The latest results \cite{Xmax} for the energy evolution of the first two moments of $X_\mathrm{max}$, superimposing FD and SD measurements, are shown in figure \ref{fig:fig1}. Their comparison with the extreme scenarios of proton and iron showers, simulated using different high-energy interaction models, reveals not only that the models generate values of $X_\mathrm{max}$ consistent with those measured. It also reveals that the average mass of cosmic rays evolves towards a lighter composition between $10^{17.2}$ and $10^{18.3}$ $\si{eV}$, which is linked to their origin \cite{Origin}. At higher energies, the trend is reversed, and the average mass increases with energy.
 
However, when comparing the energy evolution of the mean of $X_\mathrm{max}$ with that of its standard deviation, a certain tension can be observed for QGSJetII-04. For energies close to $10^{18.2}$ $\si{eV}$ and according to QGSJetII-04, the mean of $X_\mathrm{max}$ corresponds to that of a pure proton flux. At the same time, the standard deviation of $X_\mathrm{max}$ is lower than what it would be for a pure proton scenario.

\subsection{Number of muons}

\begin{figure}
\hspace{1.1cm}
\begin{minipage}{0.32\linewidth}
\centerline{\includegraphics[width=1.55\linewidth]{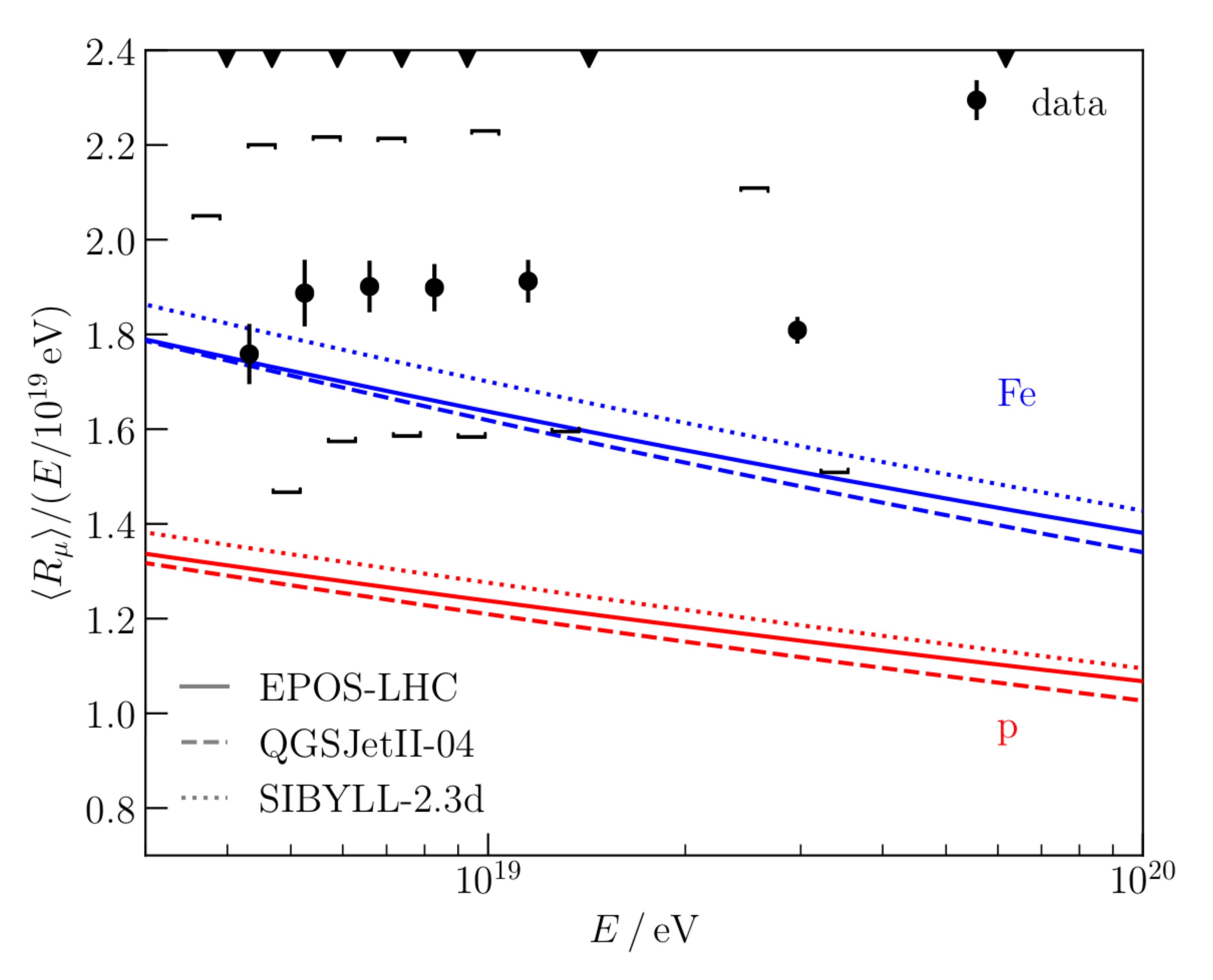}}
\end{minipage}
\hspace{2.8cm}
\begin{minipage}{0.32\linewidth}
\centerline{\includegraphics[width=1.6\linewidth]{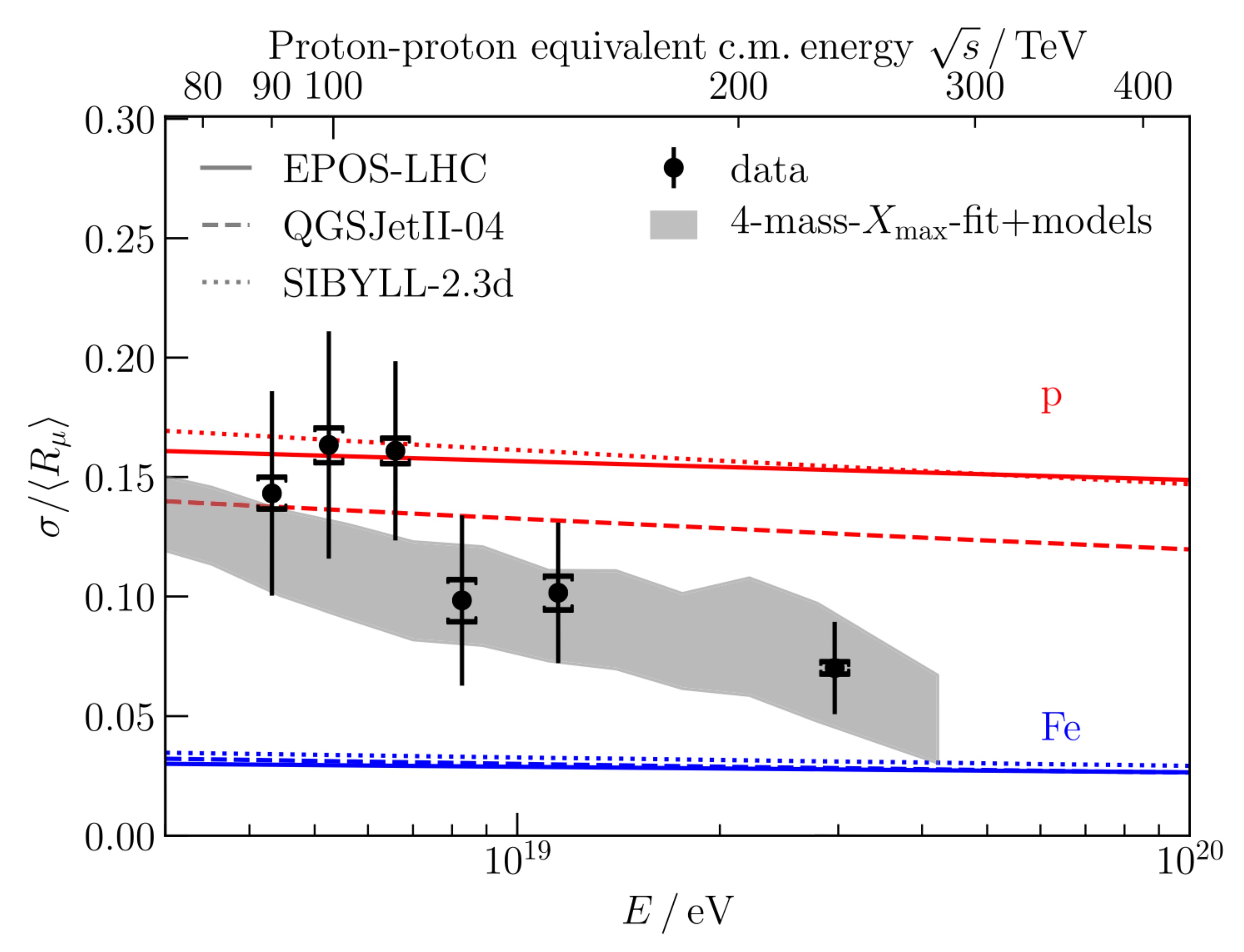}}
\end{minipage}
\caption[]{Energy evolution of the average number of muons (left) and its relative fluctuation (right) measured at the Pierre Auger Observatory for inclined showers. The results are compared to proton and iron shower simulations. Figures from \cite{Nmu, Suppl}.}
\label{fig:fig2}
\end{figure}

$\quad \;$ The number of muons is also a mass sensitive observable because the higher the mass of the primary cosmic ray, the more muons are produced. One way to obtain the number of muons from extensive air showers is by considering only inclined showers (exceeding 62$^\circ$). For these, the electromagnetic component is absorbed in the atmosphere, and the particles measured at the ground are only muons or their decay products. The average number of muons and its relative fluctuation, measured this way at the Pierre Auger Observatory \cite{Nmu}, are shown as a function of energy in figure \ref{fig:fig2}. More specifically, $R_\mu$ is the integrated number of muons at the ground divided by a reference value given by $\langle N_\mu \rangle$ for simulated showers at $10^{19}$ $\si{eV}$. The obtained values are compared to the extreme proton and iron scenarios from simulations. This time, it is evident that the mean number of muons is not well reproduced by simulations, while the measurement of the relative fluctuations falls within the range that is expected from current high-energy hadronic interaction models. Simulations, performed using any of these models, yield a number of muons which is low compared to measurements. This is the so-called muon deficit problem.

\begin{figure}
\hspace{1.3cm}
\begin{minipage}{0.32\linewidth}
\centerline{\includegraphics[width=1.75\linewidth]{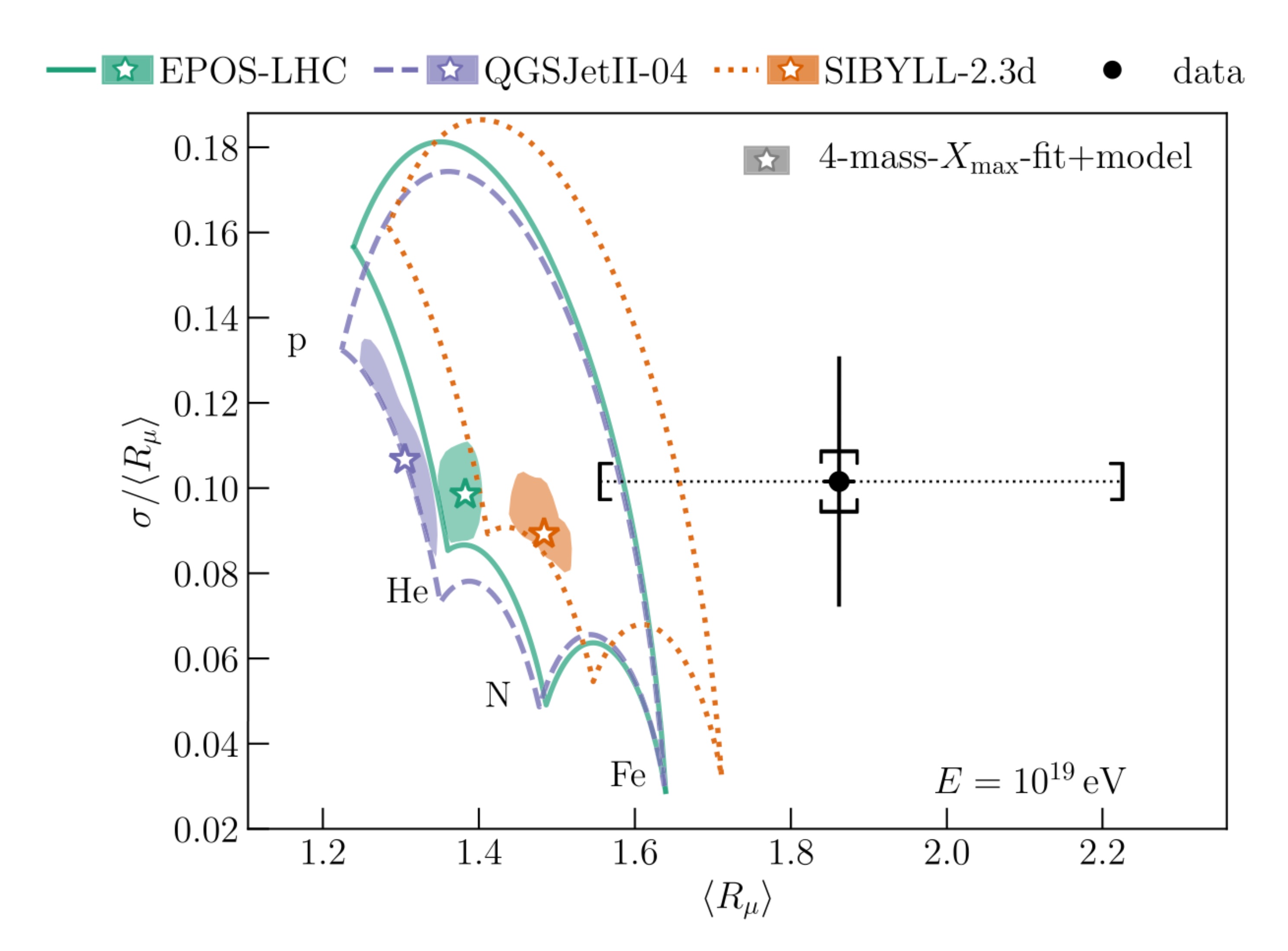}}
\end{minipage}
\hspace{3.cm}
\begin{minipage}{0.32\linewidth}
\centerline{\includegraphics[width=1.55\linewidth]{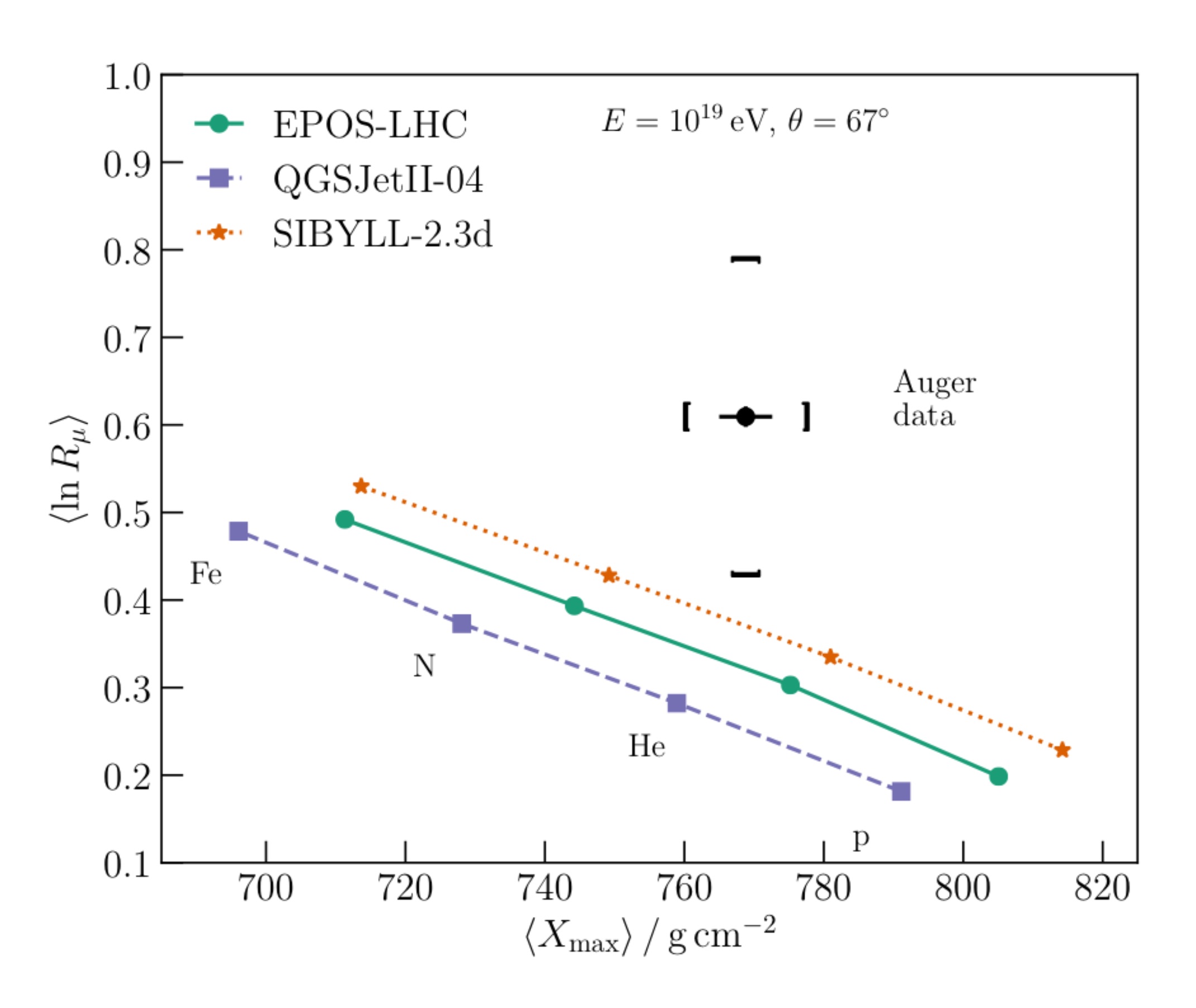}}
\end{minipage}
\caption[]{Left: Data measured at the Pierre Auger Observatory (black dot) compared to the average number of muons and its relative fluctuation obtained from simulations. The simulated expected values are obtained for any mixture of the four primaries p, He, N and Fe and are represented by the colored contours. The star symbols indicate the values that correspond to the mass composition mixture derived from $X_\mathrm{max}$ measurements. Right: Average logarithmic muon content as a function of the average maximum shower depth. Figures from \cite{Nmu, Suppl}.}
\label{fig:fig3}
\end{figure}

In order to perform a comparison between measured and simulated values of the average number of muons and its relative fluctuation simultaneously, a particular energy value needs to be fixed. The result obtained for a primary energy of $10^{19}$ $\si{eV}$ is shown in figure \ref{fig:fig3}, left. Here, not only the extreme scenarios of proton and iron initiated showers were used to construct the region of possible values covered by simulations. Sets of helium and nitrogen initiated showers and of all possible combinations of these four different primaries were considered as well (represented by colored contours). None of the predictions from the hadronic interaction models, considering a mass composition mixture derived from $X_\mathrm{max}$ measurements (represented by stars), is consistent with the measurement (represented by the black dot). The increases in the average number of muons, necessary to reconcile the simulated values with the measurement, vary from 26$\%$ to 43$\%$, depending on the high-energy interaction model used in the simulations.

The comparison between measured and simulated values of the mean logarithmic muon content and the average maximum shower depth are shown in figure \ref{fig:fig3}, right. Also here it is evident that, even though the simulated mean value of $X_\mathrm{max}$ is consistent with data, the mean logarithmic muon content is not well reproduced. The correction(s) needed in the high-energy hadronic interaction models to reproduce the muon content correctly might very well change the models in a way that the mean $X_\mathrm{max}$ values are also altered. From figure \ref{fig:fig3} (right) it becomes clear that this would lead to a change in the interpretation of the mass composition. 

%It is well known that there is a strong relationship between the number of muons and $X_\mathrm{max}$. This is related to the distribution of energy along the shower between the different components. The more energy goes into the production of neutral pions, which almost instantly decay to two photons, the more energy is diverted to the electromagnetic component. This yields a higher value of $X_\mathrm{max}$ and means at the same time that less energy is kept in the hadronic core. The latter leads to a lower number of muons, which are the decay products of hadronically interacting particles. 

When considering the measured cosmic ray flux, which implies having a mixed composition, an anti-correlation between the number of muons and $X_\mathrm{max}$ can be observed. This derives from the fact that showers from heavier primaries develop quicker in the atmosphere, rendering thus smaller $X_\mathrm{max}$-values. At the same time, this means a higher number of muons. This strong interplay between the number of muons and $X_\mathrm{max}$ motivates the study presented in the next section.

\subsection{Adjustments to the predictions of $X_\mathrm{max}$ and the hadronic signal at the ground}\label{Vicha}

$\quad \;$ Considering observables one by one only indicated that increasing the muon content without the necessity of changing the values of $X_\mathrm{max}$ was enough to obtain a consistent picture. However, performing ad-hoc adjustments to the predictions from hadronic interaction models in order to improve their consistency with observed two-dimensional distributions proves different \cite{Vicha}. In this work, two-dimensional distributions of the depth of shower maximum $X_\mathrm{max}$ and the signal at the ground were considered. In order to improve the consistency, a global shift $\Delta X_\mathrm{max}$ of the predicted shower maximum and a re-scaling $R_{\mathrm{Had}}$ of the simulated hadronic component at the ground (which includes muons and electromagnetic halos from their decays) are allowed. The latter depends on the zenith angle.

\begin{figure}
\hspace{1.4cm}
\begin{minipage}{0.32\linewidth}
\centerline{\includegraphics[width=1.72\linewidth]{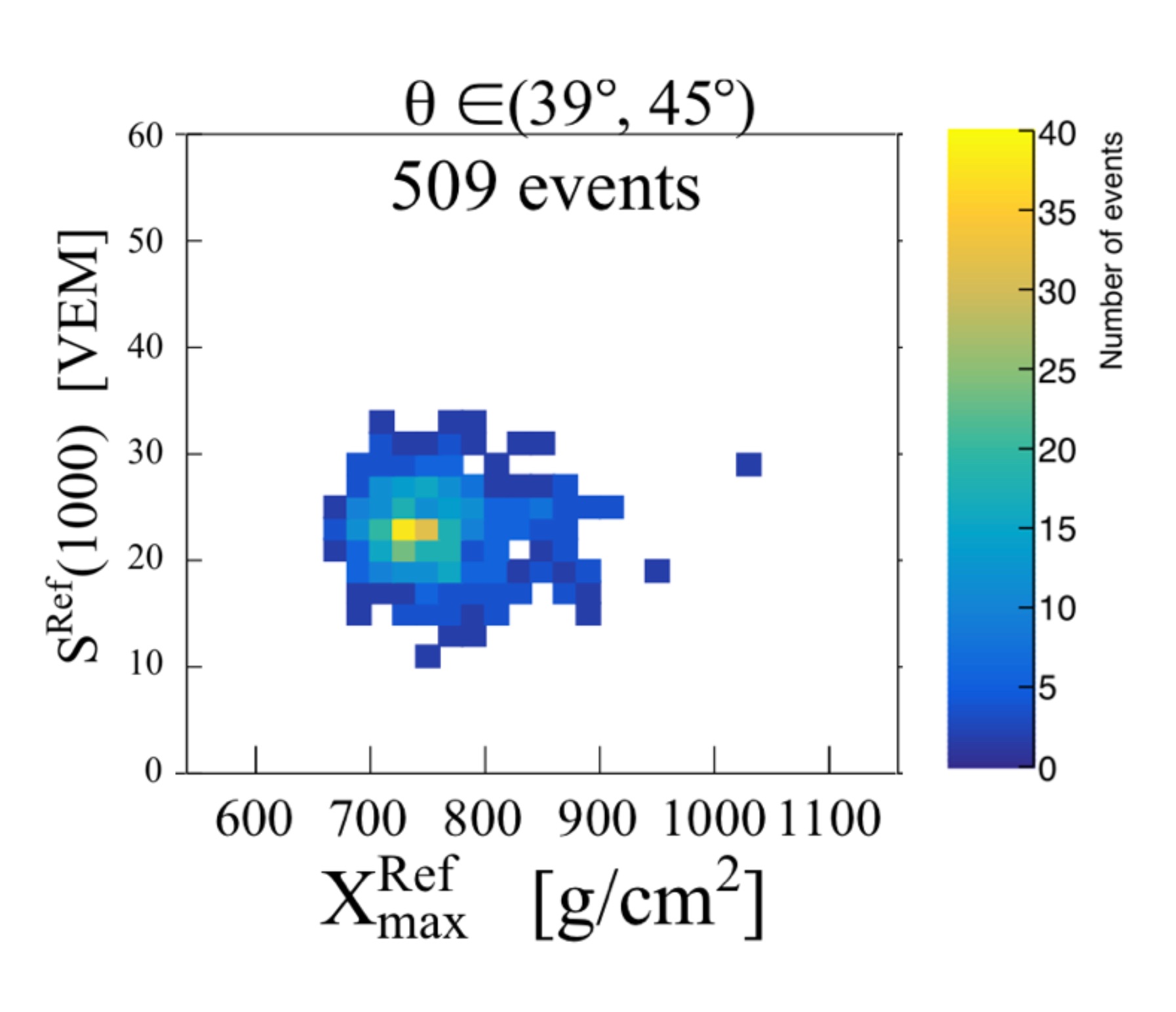}}
\end{minipage}
\hspace{3cm}
\begin{minipage}{0.32\linewidth}
\centerline{\includegraphics[width=1.57\linewidth]{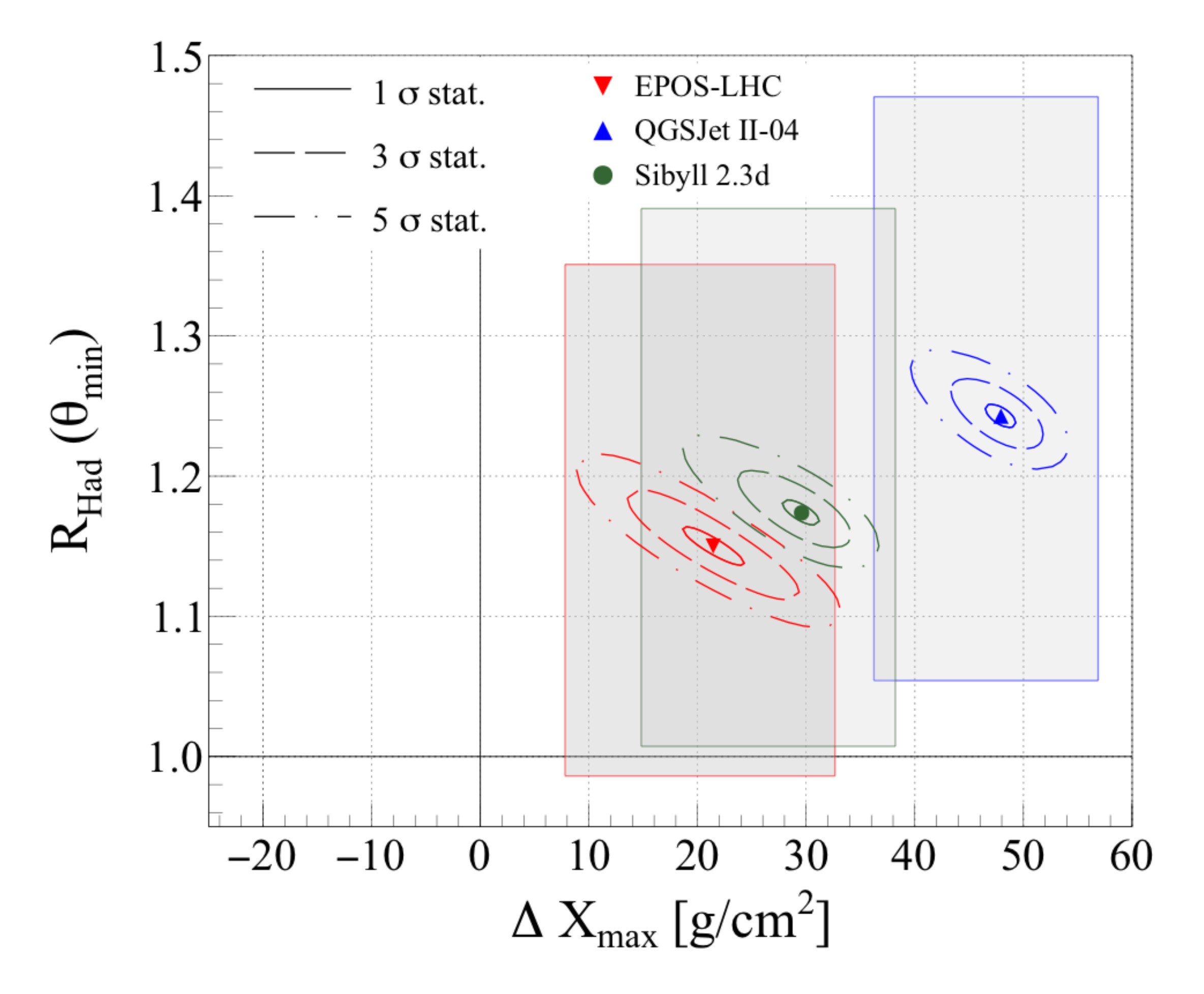}}
\end{minipage}
\caption[]{Left: Two-dimensional distributions of $X_\mathrm{max}^{\mathrm{Ref}}$ and $S^{\mathrm{Ref}}(1000)$ measured at the Pierre Auger Observatory in the energy range $10^{18.5}-10^{19.0}$ $\si{eV}$. Data are divided into five zenith angle bins. The distribution for the bin in the middle is shown here. Right: Correlation between the re-scaling parameter of the hadronic signal for the minimal zenith angle bin and the shift parameter of the depth of maximum development. Different high-energy hadronic interaction models are considered. Figures from \cite{Vicha}.}
\label{fig:fig4}
\end{figure}

The method is based on a maximum likelihood fit of the two-dimensional distributions of $X_\mathrm{max}$ and the signal at the ground at a distance of 1000 m from the core, denoted by $S(1000)$. This signal is measured by the water-Cherenkov stations of the SD at the Pierre Auger Observatory. The attenuation of this signal depends on the angle of incidence of the shower and is accounted for in this approach by the factor $f_{SD} (\theta)$. Since this attenuation is model-dependent, the data are divided into five zenith angle bins (0$^\circ$, 33$^\circ$, 39$^\circ$, 45$^\circ$, 51$^\circ$, 60$^\circ$). The distribution for the bin in the middle is shown in figure \ref{fig:fig4}, left (Ref indicates that the observables are given in reference to a fixed energy $E^{\mathrm{Ref}} = 10^{18.7}$ $\si{eV}$). These distributions are fitted with sums of templates. The sum goes over the four individual primary species considered (p, He, N, Fe) and is weighted by their relative fractions. Since these sum up to 1, the fractions serve as three free parameters. The templates are the product of a Generalized Gumbel function and a Gaussian function, which describe the $X_\mathrm{max}^{\mathrm{Ref}}$ and $S^{\mathrm{Ref}}(1000)$ distributions, respectively. The ad-hoc adjustments $\Delta X_\mathrm{max}$ and $f_{SD}$ are applied to the simulated values $\widehat{X_\mathrm{max}^{\mathrm{Ref}}}$ and $\widehat{S^{\mathrm{Ref}}}(1000)$ to obtain the corrected values
\begin{eqnarray*}
    X_\mathrm{max}^{\mathrm{Ref}} & \equiv & \widehat{X_\mathrm{max}^{\mathrm{Ref}}} + \Delta X_\mathrm{max} , \\
    S^{\mathrm{Ref}}(1000) & \equiv & \widehat{S^{\mathrm{Ref}}}(1000) \cdot f_{SD} (\theta).
\end{eqnarray*}
This adds $\Delta X_\mathrm{max}$, $R_{\mathrm{Had}}(\theta_\mathrm{min})$ and $R_{\mathrm{Had}}(\theta_\mathrm{max})$ to the set of free parameters. $f_{SD} (\theta)$ is a function of the latter two, which are the re-scaling parameters at the two extreme zenith angle bins. 

Indeed, for all three high-energy interaction models, a prediction of a deeper $X_\mathrm{max}$ is preferred, with $\Delta X_\mathrm{max}$ ranging from 22 $\si{g/cm^2}$ to 48 $\si{g/cm^2}$, depending on the model. This increase in the Monte Carlo prediction of $X_\mathrm{max}$ results in an increase of the signal at the ground due to the electromagnetic component, which alleviates the muon deficit problem present in the predictions. Increases of between 15$\%$ and 24$\%$ are needed at low zenith angles, depending on the high-energy hadronic interaction model used. For high zenith angles, the necessary increases vary from 14$\%$ to 17$\%$. These results are summarized in figure \ref{fig:fig4}, right. As can be intuited from figure \ref{fig:fig3}, right, and is presented in detail in \cite{Vicha}, the corrections on $X_\mathrm{max}$ to higher values imply a heavier mass composition compared to inferences made using the unaltered hadronic interaction models. This work shows the importance of considering a multivariate approach in order to not only obtain a better understanding, but also because this might have an impact on further conclusions.

\section{Direct approaches}

$\quad \;$ It is possible to perform direct studies with current air shower models. Typically, this involves two steps. First, using air shower simulations, a relationship between one or more observables and one or more physical properties must be established. This renders a conversion function that can be used in a second step to deduce the values of the physical parameters corresponding to the measured values of the observables.

\subsection{Proton-air cross-section}

\begin{figure}
\hspace{1.3cm}
\begin{minipage}{0.32\linewidth}
\centerline{\includegraphics[width=1.68\linewidth]{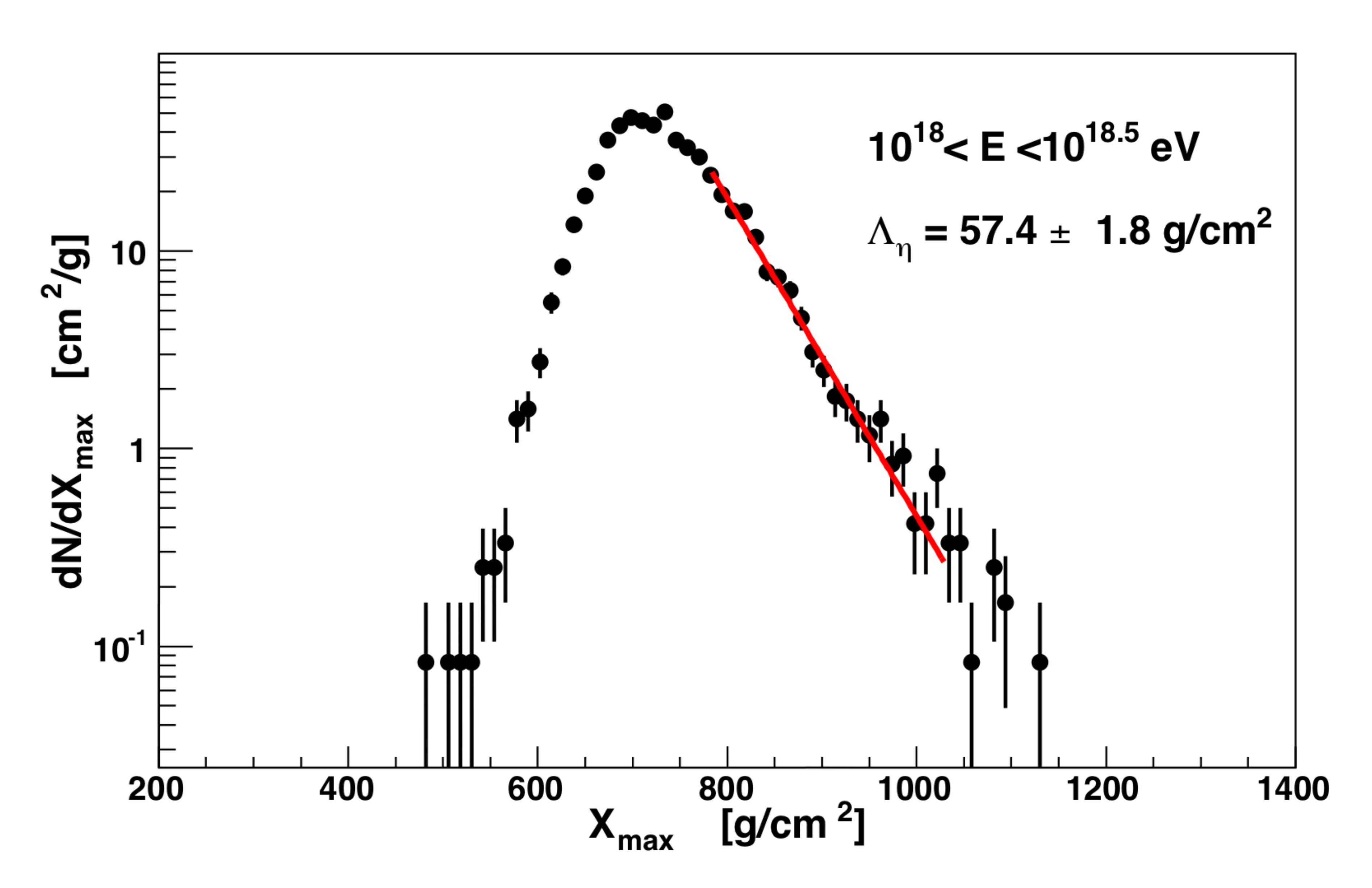}}
\end{minipage}
\hspace{2.8cm}
\begin{minipage}{0.32\linewidth}
\centerline{\includegraphics[width=1.65\linewidth]{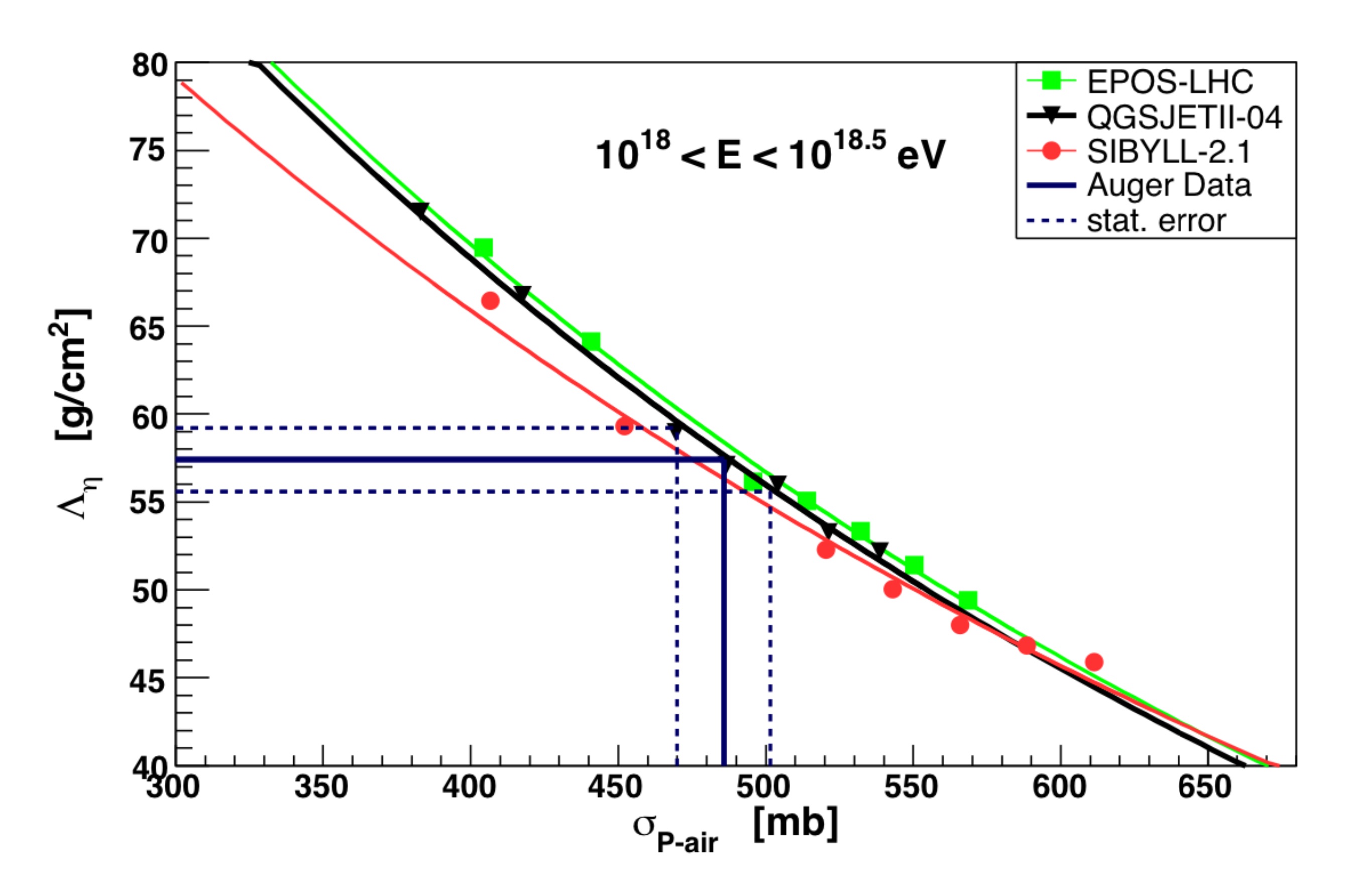}}
\end{minipage}
\caption[]{Left: $X_\mathrm{max}$-distribution in one of the studied energy intervals as measured at the Pierre Auger Observatory. The red line is the result of the unbinned log-likelihood fit performed to obtain the parameter of interest $\Lambda_\eta$, which describes the tail of the distribution. Right: Conversion function between $\Lambda_\eta$ and the proton-air cross-section for the same energy interval and for different high-energy hadronic interaction models. Figures from \cite{CS}.}
\label{fig:fig5}
\end{figure}

$\quad \;$ The depth of maximum development $X_\mathrm{max}$ is strongly correlated with the depth where the first interaction occurs, $X_0$. The distribution of the latter, for a given energy, is given by the cross-section of proton-air interactions at that energy. Hence, it is reasonable to look in the $X_\mathrm{max}$-distribution for a parameter related to the cross-section \cite{CS}. The observable, that turns out to be related to the cross-section, is a parameter $\Lambda_\eta$, which describes the exponential shape of the tail of the $X_\mathrm{max}$-distribution via $\mathrm{d}N/\mathrm{d}X_\mathrm{max} \propto \exp{(-X_\mathrm{max}/\Lambda_\eta)}$. The $X_\mathrm{max}$-distribution, measured at the Pierre Auger Observatory for one of the studied energy intervals,  is shown in figure \ref{fig:fig5}, left, together with the unbinned log-likelihood fit. The observable $\Lambda_\eta$ was measured at the Pierre Auger Observatory using an $X_\mathrm{max}$ distribution which included a fraction  $\eta \%$ of the most deeply penetrating air showers. This is done to enhance the proton fraction in the data set used.

In order to make use of the measurement, a conversion function between $\Lambda_\eta$ and the proton-air cross-section needs to be constructed. This is done by changing cross-sections in the simulations empirically. This amounts to multiplying all hadronic cross-sections by an energy-dependent factor, which is equal to 1 for energy values for which the models used in simulations agree with LHC data. Different modifications lead to a different value of the cross-section in the energy interval considered and ultimately, after a full simulation, to a different value of $X_\mathrm{max}$. The conversion functions, obtained this way for the different hadronic interaction models, are shown in figure \ref{fig:fig5}, right. The cross-section obtained from projecting the measurement from Auger varies from 497.7 $\si{mb}$ to 523.7 $\si{mb}$, depending on the high-energy hadronic interaction model considered.

\subsection{Constraining the high-energy tail of the neutral pion spectrum}

\begin{figure}
\hspace{1.2cm}
\begin{minipage}{0.32\linewidth}
\centerline{\includegraphics[width=1.35\linewidth]{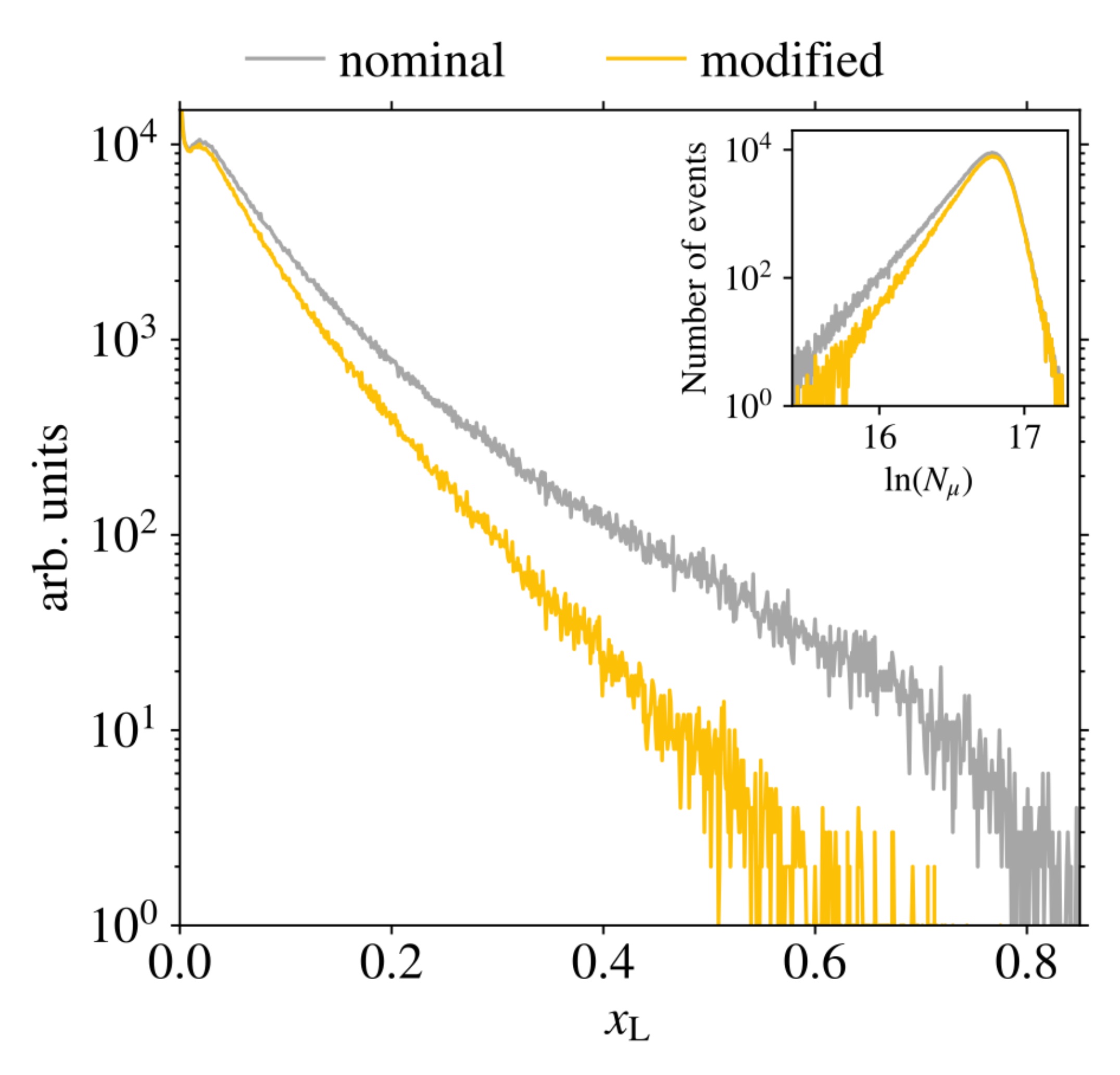}}
\end{minipage}
\hspace{3cm}
\begin{minipage}{0.32\linewidth}
\centerline{\includegraphics[width=1.55\linewidth]{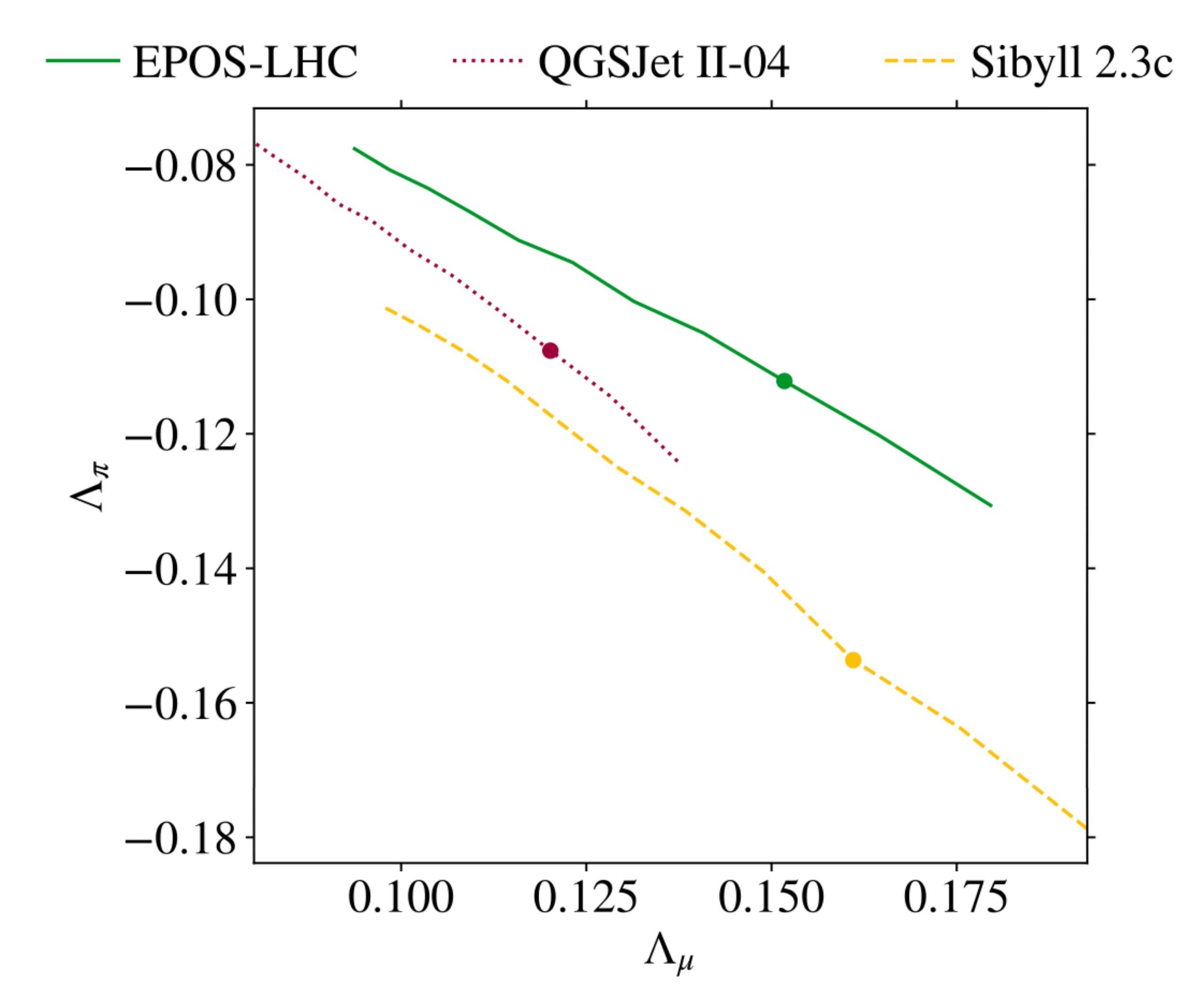}}
\end{minipage}
\caption[]{Left: Distribution of the fraction of energy carried by the most energetic $\pi^0$ in the first interaction (in the laboratory frame). The simulations are performed with proton primaries of energy $E_0= 10^{19}$ $\si{eV}$. The gray distribution corresponds to the complete set of simulations, while the yellow one corresponds to one particular re-sampling. The respective distribution of the number of muons is given in the inset plot. Right: Conversion function between the observable $\Lambda_\mu$ and the parameter $\Lambda_\pi$, which describe the low values and the tail of the corresponding distributions, respectively. The filled circles correspond to the nominal values from each hadronic interaction model. Figures from \cite{Pi0}.}
\label{fig:fig6}
\end{figure}

$\quad \;$ It is also possible to use the measured number of muons to perform a direct study on some physical property \cite{Pi0}. A lower number of muons can be attributed to an energy imbalance towards neutral pions. Since, because of isospin symmetry, the fraction of neutral pions among all pions is determined, this imbalance has to happen in the form of the energy distribution among particles. Thus, very energetic neutral pions must play an important role in obtaining a lower number of muons. Consequently, it is reasonable to search for a physical parameter in the high end of the $\pi^0$ energy spectrum and an observable among the lower values of $N_\mu$. The distribution of the fraction $x_{\mathrm{L}}$ of the energy carried by the most energetic neutral pion of the first interaction is shown in figure \ref{fig:fig6}, left (gray distribution). The distribution of the number of muons is in the inset plot. The showers considered here are proton simulations with a primary energy of $E_0= 10^{19}$ $\si{eV}$. The lower values of $\ln{(N_\mu)}$ can be described through a shape parameter $\Lambda_\mu$ by means of expression $B \exp{( \ln{(N_\mu)} / \Lambda_\mu )}$. Similarly, the tail of the $x_{\mathrm{L}}$-distribution can be fitted by the exponential function $C \exp{( x_{\mathrm{L}} / \Lambda_\pi )}$. 

It turns out that there is a relation between the physical parameter $\Lambda_\pi$ and the observable $\Lambda_\mu$ (see figure \ref{fig:fig6}, right). In this work, the conversion function is obtained by introducing a small perturbation in the $x_{\mathrm{L}}$- and $\ln{(N_\mu)}$-distributions. This is achieved by re-sampling the original distributions (gray distributions). An example of modified distributions is also given in figure \ref{fig:fig6}, left (yellow distributions). As can be seen, a small perturbation in the $\ln{(N_\mu)}$-distribution induces a small change in the $x_{\mathrm{L}}$-distribution. This results in a slightly different conversion curve for each high-energy hadronic interaction model considered. A precise determination of $\Lambda_\mu$ is still necessary, which will be possible once the number of events is large enough. In this pursuit, upgrades, such as AugerPrime, will play a crucial role.

\section{Summary and outlook}

$\quad \;$ With the Pierre Auger Observatory, interesting extensive air shower observables can be measured. On the one hand, the distribution of the depth of maximum development $X_\mathrm{max}$ and, on the other hand, the muon content were compared to the corresponding distributions obtained from simulations. Here, different available high-energy hadronic interaction models were used. It was observed that the number of muons is not well reproduced. In addition, two-dimensional distributions of $X_\mathrm{max}$ and the signal at the ground at a distance of 1000 m from the core were considered. This study revealed that, in fact, also the simulated $X_\mathrm{max}$-values deserve further investigation. Under the conditions considered in \cite{Vicha}, the corrections performed on $X_\mathrm{max}$ alleviate the muon deficit present in simulations.

It is also possible to use air shower simulations to perform direct studies. There are different techniques to introduce changes in some model parameters in order to investigate their effect on air shower observables. This way, it was observed that the steepness of the tail of the $X_\mathrm{max}$-distribution is related to the proton-air cross-section. This allowed the measurement of this cross-section at energies beyond those achievable at the LHC. Similarly, once enough events are recorded at the Pierre Auger Observatory, it will be possible to use measurements of the number of muons to obtain constraints on the high-energy tail of the neutral pion spectrum. 

It is worth emphasizing that performing a bi-variate comparison between simulations and measurements (see section \ref{Vicha}) revealed inconsistencies not evident from one-dimensional studies. With this in mind, it makes sense to think that also multivariate direct approaches might give access to physical properties not accessible with only one observable. 

\section*{Acknowledgments}
$\quad \;$ I thank Tanguy Pierog for the fruitful discussions and the organizers of the Rencontres de Moriond for inviting me to contribute with this talk. I am supported by LabEx UnivEarthS (ANR-10-LABX-0023 and ANR-18-IDEX-0001).

\end{document}